\documentclass[
    ,final            % use final for the camera ready runs
  ]
  {aipproc}

\layoutstyle{6x9}

\begin{document}

  \title{The Earth's Gamma-ray Albedo as observed by EGRET}

\author{Dirk Petry}{
  address={Joint Center for Astrophysics, University of Maryland Baltimore 
    County, Baltimore, MD 21250\\
 and NASA, Goddard Space Flight Center, Greenbelt, MD 20771}
}

\begin{abstract}
  The Earth's high energy gamma-ray emission is caused by cosmic ray
  interactions with the atmosphere. The EGRET detector on-board the CGRO
  satellite is only the second experiment (after SAS-2) to provide a 
  suitable dataset for the comprehensive study of this emission. 
  Approximately 60\% 
  of the EGRET dataset consist of gamma photons from the Earth.
  This conference contribution presents the first results from the
  first analysis project to tackle this large dataset. Ultimate purpose
  is to develop an analytical model of the Earth's emission for 
  use in the GLAST project. 
  The results obtained so far confirm the earlier results 
  from SAS-2 and extend them in terms of statistical precision and 
  angular resolution.  
  
\end{abstract}

\maketitle

%%%%%%%%%%%%%%%%%%%%%%%%%%%%%%%%%%%%%%%%%%%%
%% MAINMATTER
%%%%%%%%%%%%%%%%%%%%%%%%%%%%%%%%%%%%%%%%%%%%

\section{Motivation}

The Earth's high-energy gamma-ray emission (the ``gamma-ray albedo'') 
has been studied very little from space. The most precise measurements so far
were obtained in 1972/73 with the SAS-2 satellite \cite{thompson} but have very limited
photon statistics. Other space measurements include \cite{kraushaar,guryan,galper}.

The gamma-ray albedo is caused by the interaction of cosmic rays and solar
wind with the Earth's atmosphere.
The study of the albedo is interesting mainly for two reasons: 

\begin{enumerate}
\item From a geophysical point of view, it can serve to improve our 
understanding of the interaction of cosmic rays and solar wind with the atmosphere and
the Earth's magnetic field.

\item For gamma-ray-astronomical observations from satellites, the albedo constitutes an intense background.
It is important to know the properties of this background in detail in order to
be able to 
(a) avoid it as much as possible by specific pointing strategies and
data filtering (either online or offline),  and 
(b) possibly use it as a  calibration source.
\end{enumerate}

\noindent
The EGRET high-energy gamma-ray telescope \cite{thompson2} was operational 
as part of the Compton Gamma-Ray
Observatory  satellite (CGRO) for 9.1 years from 1991 until 2000.
CGRO had a circular orbit at approx. 450 km altitude (the altitude changed over time)
with an inclination of 28.5$^\circ$.
From this perspective, the Earth's horizon appears at a zenith angle of $\approx$111$^\circ$.
The instrument was almost always in pointed observation mode which meant that the Earth
occulted the object of interest for on average nearly 50\%  of the time, or in other words,
the Earth was in the 80$^\circ$ diameter field of view (FOV) for more than half the time.

In order to avoid the bright gamma-ray albedo of the Earth, the spark-chamber trigger of
EGRET was automatically modified as the Earths horizon moved in and out of the FOV. 
Still, the number of albedo photons registered by EGRET is large.
The total cleaned photon dataset comprises 5.2 million events. Roughly 60\% of these were observed
from zenith angles $> 105^\circ$. More than half of the archival EGRET data are albedo photons!

In view of the upcoming GLAST mission \cite{michelson}, the interest in the Earth's albedo
increased, and it was decided to use the rich EGRET dataset to obtain the most detailed
picture of the Earth's albedo available so far. The application of this knowledge to GLAST
will be especially direct because GLAST is foreseen to fly an orbit very
similar to that of CGRO.

The results presented here are from an ongoing analysis project and 
make use of only about 10\% of the available data. In this first step, the analysis
method was established and results of previous observations confirmed.
The project will continue to study the complete data set, and look in addition
at longitudinal and latitudinal variations of the measured parameters.
 
\section{Method}

All results were obtained using the newly developed software DAVE \cite{dave} (written 
in C++ and based on tools under development for the GLAST mission).
The EGRET pointing and livetime history was converted into
an Earth-centered coordinate system interpolating between entries
on a one-minute timescale.
Taking into account the changing detector modes and assuming a differential gamma-ray
spectral index of 2.0 (the value measured by SAS-2 for the horizon emission), 
all-sky exposure maps in a Hammer-Aitoff projection were obtained. Using the same 
Earth-centered coordinate system,
corresponding counts maps were derived from the event data. Finally, taking into account the
solid angle of each map pixel, flux maps were obtained. This process was repeated at different
angular resolutions, in different energy bands, and restricting
the EGRET data to different Earth latitude bands. All error calculations used
the appropriate Poisson statistics.

\section{Results}

The results shown here constitute the first installment of the first ever
Earth gamma-ray albedo study using EGRET data. 

The main aim in the first phase of this analysis project was to produce results which
can be compared to the earlier results by SAS-2 \cite{thompson}. 
SAS-2 had an equatorial orbit with only 2$^\circ$ inclination (as opposed to 28.5$^\circ$
for EGRET).
To enable the comparison, the satellite latitude in the EGRET dataset was
restricted to the range $\pm 5^\circ$.
Due to the proximity to the peak of the Solar cycle, CGRO was exposed to strong
atmospheric drag during the first mission years such that its altitude decreased relatively
quickly. In order to have a constant definition of the zenith angle, the data
was restricted to the first year of EGRET operation
(viewing periods 0.2 until 25.0).

Using the method described above, flux maps were obtained with 1$^\circ$ resolution
for the energy bands \ \  35\,MeV \-- 100\,MeV,\ \  
100\,MeV \-- 300\,MeV, 300\,MeV \-- 1\,GeV, and 1\,GeV \-- 10\,GeV.
In these maps, especially at higher energies, some pixels stay empty due to low photon
statistics. 

\begin{figure}
  \includegraphics[width=\textwidth]{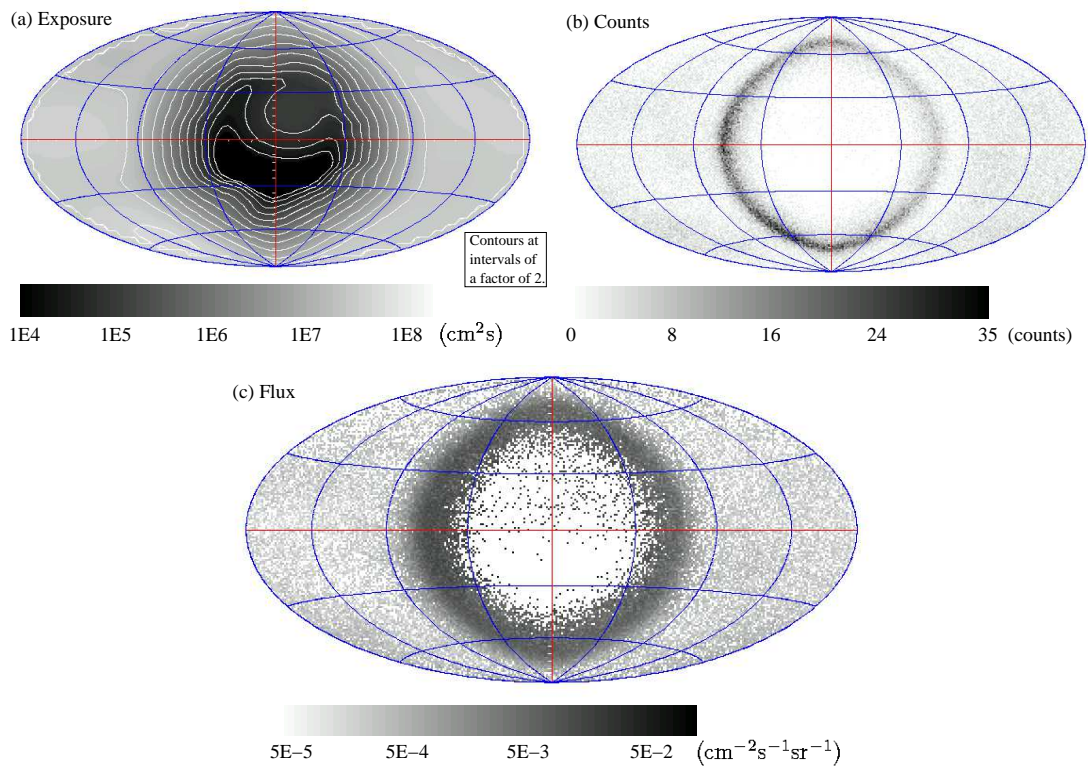}
  \caption{\label{fig-maps}
  Examples of Earth-centered Hammer-Aitoff all-sky maps derived from the first year of EGRET data
  in the energy range 100 MeV - 300 MeV at geographical satellite latitudes
  in the range $\pm 5^\circ$ 
  (North is up, East is to the right, pixel size $\approx$ 1$^\circ\times$1$^\circ$): (a) Exposure map, 
  (b) Counts map, (c) Flux map = (b) divided by (a) times the pixel solid
  angle map.}
\end{figure}

\begin{figure}
  \includegraphics[width=\textwidth]{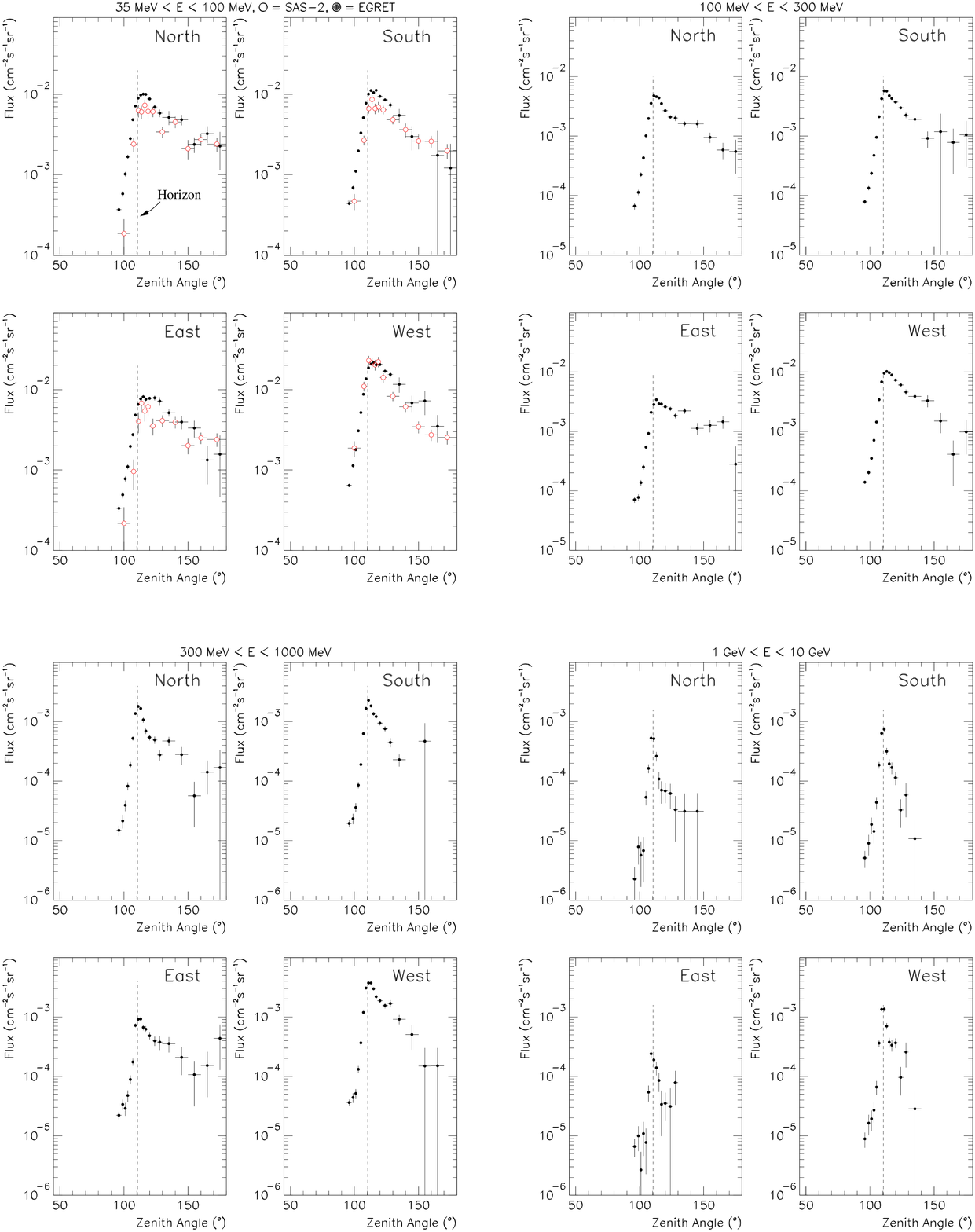}
  \caption{\label{fig-radial} 
  Radial flux profiles derived from the same
  dataset as figure \ref{fig-maps} for the four sectors North, South, East, and
  West in the four energy bands defined in the text . 
  In the plot for the first energy band, the published results from SAS-2 are 
  shown for comparison.}
\end{figure}

\begin{figure}
  \includegraphics[width=\textwidth]{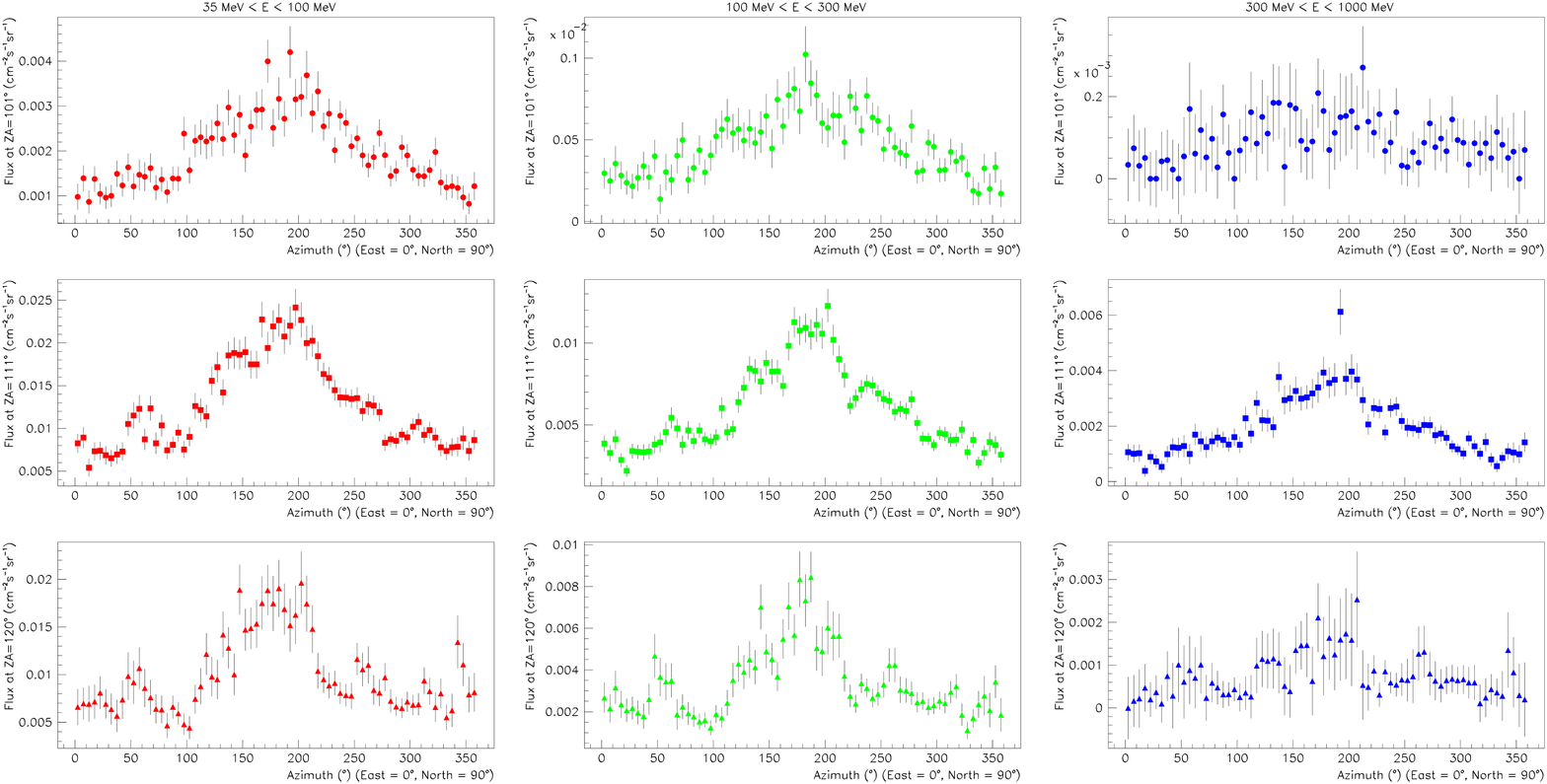}
  \caption{\label{fig-azimuthal}
  Azimuthal flux profiles for three 4$^\circ$ zenith angle bins
  around 101$^\circ$, 111$^\circ$ (i.e. the horizon), and 120$^\circ$ derived
  from the same dataset as figures \protect\ref{fig-maps} and \protect\ref{fig-radial}.}
\end{figure}

From the 1$^\circ$ resolution maps, azimuthal and radial (i.e. along lines of constant azimuth)
flux profiles were obtained, the latter in four 60$^\circ$ sectors: North (azimuth 60$^\circ$ \-- 120$^\circ$),
South (azimuth 240$^\circ$ \-- 300$^\circ$), East (azimuth 330$^\circ$ \--
30$^\circ$), West (azimuth 150$^\circ$ \-- 210$^\circ$). Figures
\ref{fig-radial} and \ref{fig-azimuthal} show the profiles.

\subsection{Comparison to SAS-2 data}
EGRET confirms the SAS-2 results (see figures
\ref{fig-radial} and \ref{fig-spec}). 
The comparison is meaningful also because the observations were made with a time
separation of 19 years, i.e. with a phase shift of only 0.2 in solar cycle. 
% cycle is 10.5 years this century
However, the number of sunspots in 1973 was significantly lower than in 1992.
Furthermore, SAS-2 had a more eccentric orbit with the altitude varying between 440\,km and 610\,km. 

\subsection{Radial Profiles}
The radial flux profiles can be described by a nearly Gaussian edge towards smaller zenith angles,
a peak very near the geometrical horizon position, and a more complex, nearly
exponential decay towards larger zenith angles.

The half width of the Gaussian edge is decreasing with increasing energy
by about 2$^\circ$ per decade in energy (taking into account corrections for the energy
dependence of the point spread function) while the peak position converges
towards the horizon (110.4$^\circ$) starting at about 115$^\circ$ at 60 MeV.
At about 3 GeV, the peak position becomes indistinguishable from the horizon.

Seen from the Equator, the North and South sectors show no significant difference.
The difference between East and West, however, is large. This is expected from the interaction of
the cosmic rays with the Earth's magnetic field and was already observed by previous experiments.
The West/East peak ratio increases with energy from approx. 2.5 at 60\,MeV to approx.
5.5 at 3\,GeV.

\subsection{Azimuthal Profiles}
The azimuthal flux intensity profile at the horizon shows the Western intensity peak to be very
broad with a half width of about 45$^\circ$. This feature does not depend significantly on energy.

\subsection{Spectra}
\begin{figure}
  \includegraphics[width=0.7\textwidth]{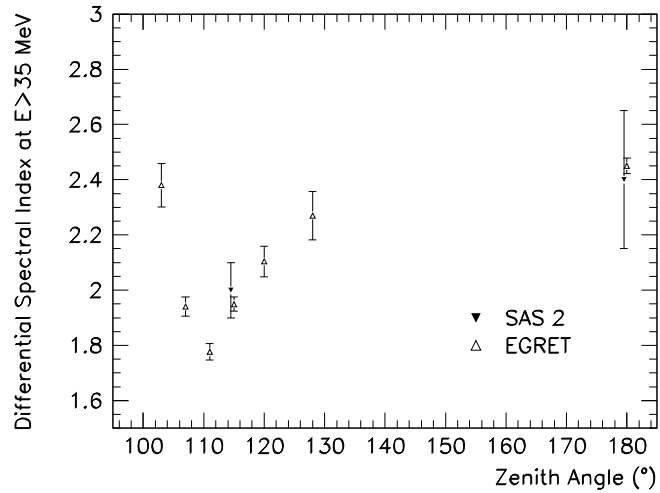}
  \caption{\label{fig-spec}
   The results of power-law fits to the spectra between 60\,MeV and 3\,GeV (energy
   bin centers) 
   averaged over the total azimuth range at different zenith angles, plotted
   vs. the zenith angle. In addition, the published SAS-2 results are shown
   for comparison.}  
\end{figure}

The gamma-ray spectrum between 60 MeV and 3 GeV (energy bin centers) 
is generally reasonably well described by a power law.
It is hardest (index 1.76$\pm$0.03$\pm$0.05$_{syst}$) 
at the horizon peak position  and softer both outside and inside
(see figure \ref{fig-spec}). The variation of the spectrum between the four
60$^\circ$ sectors (North, South, East, and West) is not significant in this dataset.

\begin{theacknowledgments}

The author would like to thank B. Dingus for suggesting this project, 
D.L. Bertsch, J. Chiang, and S. Bansal for contributions
to the analysis software, and D.J. Thompson, 
D.L. Bertsch, and R.C. Hartman for useful discussions. 
\end{theacknowledgments}

\end{document}